\documentclass[a4paper,11pt]{article}

\newcommand\s{\mathrm{s}}  

\newcommand\cm{\mathrm{cm}} 
\newcommand\km{\mathrm{km}} 

\newcommand\g{\mathrm{g}}   

\newcommand\erg{\mathrm{erg}} 

\newcommand\ergsec{\erg/\s}  

\newcommand\sh{\mathrm{sh}} 
\newcommand{\mach}{\mathcal{M}} 
\newcommand{\bo}{\mathrm{bo}} 

\newcommand{\choke}{\mathrm{ch}} 
\newcommand{\engine}{\mathrm{e}} 
\newcommand{\te}{t_\mathrm{e}} 
\newcommand{\jet}{\mathrm{j}} 
\newcommand\head{\mathrm{h}} 

\newcommand{\thresh}{\mathrm{thresh}} 
\newcommand{\sound}{\mathrm{sound}} 
\newcommand{\up}{\mathrm{u}} 

\newcommand{\de}{\mathrm{d}} 
\renewcommand{\vec}[1]{\boldsymbol{\mathbf{#1}}} 
\renewcommand{\epsilon}{\varepsilon}  

\usepackage{pos}
\usepackage{lipsum} 
\usepackage{etoolbox}
\usepackage{subcaption}
\usepackage{enumitem}
\usepackage{lineno}
\patchcmd{\thebibliography}
  {\labelsep}
  {\itemsep 0pt \labelsep}
  {}{}
  
\title{Choked jets in BSG/RSG as possible sources of high-energy neutrinos}
\ShortTitle{Choked jets in BSG/RSG as possible sources of high-energy neutrinos}

\author*[a]{Matteo Pais}
\author[b]{Angela Zegarelli}
\author[c,d]{Silvia Celli}
\author[e]{Enrico Peretti}

\emailAdd{matteo.pais@inaf.it}
\emailAdd{angela.zegarelli@astro.ruhr-uni-bochum.de}
\emailAdd{silvia.celli@roma1.infn.it}
\emailAdd{enrico.peretti.science@gmail.com}

\affiliation[a]{INAF, Istituto Nazionale di Astrofisica, Osservatorio Astronomico di Padova, Vicolo dell'Osservatorio 5, I-35122, Padova, Italy}
\affiliation[b]{Ruhr University Bochum, Faculty of Physics and Astronomy, Astronomical Institute (AIRUB), Universit\"{a}tsstra\ss e 150, 44801, Bochum, Germany}
\affiliation[c]{INFN, Istituto Nazionale di Fisica Nucleare, Sezione di Roma, P. le Aldo Moro 2, I-00185, Roma, Italy}
\affiliation[d]{Dipartimento di Fisica, Università La Sapienza, P. le Aldo Moro 2, I-00185, Roma, Italy}
\affiliation[e]{INAF, Istituto Nazionale di Astrofisica, Osservatorio Astronomico di Arcetri, Largo E. Fermi 5, 50125 Florence, Italy}

\abstract{
The death of massive stars is accompanied by the formation of central and accreting compact objects and the subsequent launch of relativistic jets. 
However, not all jets successfully drill their way out of the stellar envelope.
Unsuccessful jets, also known as choked jets, may still produce radiation at lower frequencies by dissipating the jet energy into a pressurized cocoon. This cocoon expands within the stellar envelope and eventually breaks out as a mildly relativistic outflow.
We investigate the plasma physics in the surroundings of massive collapsing stars harboring choked jets via relativistic, non-resistive MHD simulations. As a result, we define the parameter space allowing for jets to remain choked, and we quantify the acceleration rate and efficiency for charged particles in the strong shocks of such astrophysical environments. Preliminary results show that high Mach numbers ($\sim 100$) after 5-10 seconds of constant energy injection characterize the forward shock, possibly allowing for efficient particle acceleration and high-energy neutrino production.
Our results are presented for blue supergiant progenitors.
}

\ConferenceLogo{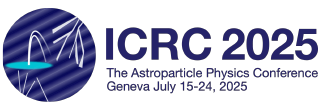}

\FullConference{39th International Cosmic Ray Conference (ICRC2025)\\
 15–24 July 2025\\
Geneva, Switzerland\\}

\begin{document}

\maketitle
\section{Introduction}
Gamma-ray bursts (GRBs) represent some of the most energetic and transient phenomena in the cosmos, typically linked to catastrophic events such as the mergers of compact objects or the collapse of massive stars. In particular, the death of massive stars, with initial masses exceeding $\approx 8~M_\odot$, can trigger core-collapse supernovae (CCSNe), leading to the formation of a compact remnant such as a neutron star or black hole. If sufficient material accretes onto this remnant, a powerful relativistic jet may be launched \citep{gottlieb_black_2022}. When the progenitor has been stripped of its outer layers, the collapse often results in a Type Ib/c supernova (SN) and, in some cases, a successful long GRB. However, if the star retains its hydrogen-rich envelope, as in the case of Red and Blue Supergiants (RSGs and BSGs), the jet may be unable to escape, resulting in what is referred to as a choked GRB.

Although these failed jets are suppressed in gamma-ray emission due to the high optical depth of the dense stellar envelope, they remain of significant astrophysical interest. 
As the jet propagates through the star, it drives a high-pressure cocoon at its head, redirecting energy sideways \cite{bromberg_propagation_2011}.
Even if the jet stalls before reaching the surface, the associated cocoon-driven shock may still emerge from the star, producing a shock breakout signature. 
This breakout can generate a brief X-ray or ultraviolet (UV) flash, followed by days-long UV and optical transients that can serve as observable electromagnetic counterparts if the shock successfully escapes the stellar surface \cite{Piran2019, zegarelli_towards_2024}.

In parallel, choked GRBs are compelling candidates for producing high-energy neutrinos. 
The dense stellar environment facilitates efficient hadronic interactions in the jet before it is terminated, leading to the production of TeV–PeV neutrinos. 
These neutrinos can escape the progenitor with minimal attenuation, making choked GRBs promising examples of “hidden” neutrino sources, e.g., sources that are opaque to gamma rays and therefore undetectable by instruments such as those onboard Fermi \cite{Ackermann2015, murase2016}. 
This hidden-channel scenario is consistent with the lack of observed neutrino–GRB correlations reported by IceCube and ANTARES \cite{icgrb, antares2021}, as well as with the unknown origin of the diffuse astrophysical neutrino flux detected by IceCube since 2013 \cite{icecube_diffuse}.

Given their potential to accelerate ultra-high-energy cosmic rays (UHECRs) \cite{baerwald2015} and to produce high-energy neutrinos \cite{waxman1997, meszaros_tev_2001}, it is crucial to understand the physical conditions that determine whether jets launched in CCSNe are successfully launched and break out, or instead become choked within the star. 
In this work, we focus on massive progenitors with extended envelopes, specifically BSGs and RSGs, as these are expected to provide favorable environments for jet choking \cite{fasano2021, He2018}. To this end, we perform dedicated relativistic magnetohydrodynamic (RMHD) simulations to study jet propagation, shock formation, and cocoon dynamics. Strong shocks in these systems are of particular interest, as they may serve as possible sites for particle acceleration.
The proceeding is structured as follows. 
In Section~\ref{sec: setup} we describe our computational setup, the jet launching mechanism, the progenitor's structure, and the initial estimates for the jet choking and engine times. 
Furthermore, we describe the shock-capturing scheme we developed, which will be applied in our post-processing analysis of our simulation. 
Finally, in Section~\ref{sec: results}, we report and discuss preliminary results on our numerical explorations of a jet launched and subsequently choked within a stellar atmosphere. We report post-processing analysis of the simulation output, including the localization of shock surfaces at different stages and the Mach Number associated with the shocked environment.

\section{Setup}
\label{sec: setup}
In the following, we describe the simulation setup for the progenitor star and the launch of the relativistic magnetized jet, as well as the criterion adopted to determine the choked-jet condition.

\subsection{Stellar setup}
For a BSG progenitor with radius $R_*$ and mass $M_*$, we take a continuous density profile given by a combination of a polytrope with a power-law in terms of $r/R_*$ for the inner profile and a power law in $r-R_*$ for the outer edge. This translates to:
\begin{equation}
\label{eq: stellar_density_profile}
    \rho(r) = \rho_0 \left( \dfrac{r}{R_*} \right)^{-\alpha} \left( 1- \dfrac{r}{R_*}\right)^n \quad , 
\end{equation}
which, integrated, gives the following density scale for $\rho_0$ such that we obtain the mass $M_*$ by integrating the density profile.
For both the BSG and RSG, we chose $\alpha=2.5$ \cite{wongwathanarat_production_2017,nakar_relativistic_2012}, 
while we chose $n=3$ for BSG and $n=3/2$ for RSG \cite{matzner_expulsion_1999, rabinak_early_2011}. 
We obtain the following estimations for $\rho_0$:
\begin{equation}
\dfrac{\rho_0}{[\g~\cm^{-3}]} = 
    \begin{cases}
        \dfrac{35 M_*}{128\pi R_*^3} = 8.14\times 10^{-3} R^{-3}_{*,50\odot} M_{*,15\odot} & \text{(BSG)} \ ,\\
        \dfrac{2 M_*}{3 \pi^2 R_*^3} = 4.79\times 10^{-8} R^{-3}_{*,500\odot} M_{*,15\odot} & \text{(RSG)} ,
    \end{cases}
\end{equation}
where radii and masses are expressed as multiples of solar values, marked by the subscript $\odot$.

\subsection{Computational setup}
We perform our simulations using the modular massively parallel multidimensional relativistic magneto-hydrodynamic code {\textsc{pluto}} (v4.4.3)\footnote{This is the latest version available during the making of this paper.} \cite{mignone_pluto_2007}. 
The code uses a finite-volume, shock-capturing scheme that integrates a system of conservation laws. The flow quantities are discretized on a rectangular computational grid enclosed by a boundary. 
We use the special relativistic hydrodynamics module in 2.5D spherical coordinates (i.e., a 2D grid in spherical geometry, i.e. $r,\theta$, in which $v_\phi$ and $B_\phi$ are also integrated and evolved with periodic boundary conditions).
We choose a Taub-Matthews equation of state \cite{mignone_equation_2007}, which accounts for both relativistic and non-relativistic fluids, with $\gamma = 5/3$ for non-relativistic material and $\gamma = 4/3$ for relativistically moving matter. 
For the magnetic fields, we employ a simple divergence cleaning scheme \cite{dedner_hyperbolic_2002}. 

\subsection{Estimating breakout and choking times}
A simple criterion to determine whether a jet is successful or not in drilling its way out of the star and blanketing it is given by the following integral equation, whose root determines the jet choking time $t_\choke$ \cite{nakar_unified_2015, pais_velocity_2023}:
\begin{equation}
\label{eq: chocking_equation}
    t - \int_0^t\beta_\head (t') \de t' = t - \dfrac{r_\head(t)}{c} =  t_\engine \quad , 
\end{equation}
where $t_\engine$ is the engine duration of the jet, and $\beta_\head(t)$ and $r_\head(t)$ are the jet head velocity and position, respectively. In this formula, we suppose that the latest element (the `jet tail') launched by the engine propagates at roughly light speed, i.e., up to a distance $z_\mathrm{t} \simeq c(t-t_\engine)$ in a time $t$. When the jet tail and the jet head become causally connected, the jet-cocoon system propagates adiabatically into the atmosphere, following the decreasing density gradient. 
Solving Eq.~\eqref{eq: chocking_equation} for $t$, an approximate expression for the choking time $t_\choke$ is given by:
\begin{equation}
\label{eq: tchoke_alpha2}
    t_\choke = \dfrac{t_\engine}{1-\langle \beta_\head \rangle} \quad .
\end{equation}
where $\langle \beta_\head \rangle$ is the time-averaged head velocity inside the star.

\begin{figure}
    \centering
    \includegraphics[width=0.5\linewidth]{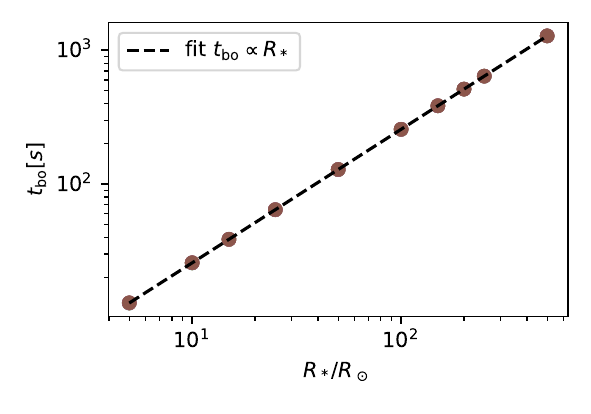}
    \caption{Breakout times of 39000 runs of our numerically-integrated 1D jet-head model where we varied the jet luminosity, jet opening angle, radius, and mass of the progenitor BSG star with a density profile with $\alpha=2.5$ and $n=3$. The points perfectly accommodate a linear fit.}
    \label{fig:linear_fit_breakout}
\end{figure}

Before running a full set of RMHD simulations, to estimate the breakout times and the choking times of the jets in our set of simulations, we decided to numerically integrate the jet's equation of motion following the methods described in \cite{hamidani_jet_2021}, which takes into account the numerical correction factor $N_s$ for the jet propagation, already introduced in \cite{harrison_numerically_2018} arisen by comparing the purely analytical results of \cite{bromberg_propagation_2011} with numerical RHD simulations. 
We run our numerical integration by giving the jet an initial radius of $r_0 = 10^4~\km$ and a cocoon radius of $r_\mathrm{c} = 0$.
We vary the jet ($L_\jet$, $\theta_\jet$) and progenitor parameters ($M_*$, $R_*$) for a total of $39000$ 1D simulations. 
Comparing and normalizing the models, and measuring their breakout times and their choking times (by root-finding the time at which the jet tail catches the jet head, assuming a quasi-luminal motion for the jet tail), we find linear scaling relations.
In Fig.~\ref{fig:linear_fit_breakout} we report the breakout time $t_\bo$ as a function of the progenitor's radius $R_*$ for our 39000 runs, obtaining a linear relation between the two.
For a progenitor with a density profile given by Eq.~\eqref{eq: stellar_density_profile} and $\alpha=2.5$, and supposing a constant and continuous supply of energy behind the jet head, the jet head at the breakout is fast enough ($(\Gamma\beta)_\head >1$) to have the stellar radius as the dominant factor, such that the breokout time is found to be:
\begin{equation}
\label{eq: tbo_radius}
   t_\bo[\s]  \simeq 10 + 116 \left( \frac{R_*}{50 R_{\odot}} \right) + 12 \left( \frac{\theta_{\rm j}}{5^{\circ}} \right)^{6/5} \left( \frac{L_{\rm j}}{10^{51}\mathrm{erg/s}} \right)^{-3/10}
\end{equation}
where $116 (R_*/50 R_\odot)$ is the breakout time of a photon launched at the same time as the jet.
If the engine time is sufficiently long, i.e. $t_\engine \gtrsim 10 \s$, implying a relativistic jet, then we find for the choking time:
\begin{equation}
\label{eq: tchoke}
    t_\choke [\s] \simeq \eta\left[ t_{\rm e}[\mathrm{s}] + 1.563 \left( \frac{t_{\rm e}}{1~\mathrm{s}} \right)^{9/7} \left( \frac{\theta_{\rm j}}{5^{\circ}} \right)^{-4/3} \left( \frac{L_{\rm j}}{10^{51}~\mathrm{erg/s}} \right)^{9/26}\right]
\end{equation}
where the engine time is expressed in seconds, and $\eta \sim 3-4$ arising from numerical uncertainty.  

Furthermore, comparing the choking to the breakout time (Eqs.~\eqref{eq: tchoke} and \eqref{eq: tbo_radius}), we can find an upper limit for the engine activity, given roughly by:
\begin{equation}
    t_\engine[\s] \lesssim \eta^{-1} \dfrac{12 L^{-3/10}_{\jet,51} \theta^{6/5}_{\jet,5^\circ} + 116 R_{50\odot} +10}{1+1.6 ~ \theta^{-4/5}_{\jet,5^\circ} L^{9/26}_{\jet,51}}
\end{equation}
where we made the dependency of the choking time linear with respect to the engine time. 
These considerations have been discussed for relativistic unmagnetized jets. 
However, the result is also valid in the case of magnetized jets, since we can absorb the contribution of the Poynting vector to the total luminosity by substituting $L_\jet$ with $L_\jet = L_\mathrm{HD} + L_\mathrm{MHD}$ which results in $\tilde{L} \rightarrow \tilde{L} (1+\sigma_\jet)$, where $\sigma_\jet$ is the jet's magnetization.

\vspace{-3mm}
\subsection{Jet injection}
The jet is characterized by five parameters: luminosity $L_\jet$, opening angle $\theta_\jet$, enthalpy $h_\jet$, magnetization $\sigma_\jet$, and engine duration $\te$. 
In our setup, we inject the jet spherically through a small polar cap of width $\theta_0 \ll \theta_\jet$ such that the initial Lorentz factor $\Gamma_{0,\jet}$ is related to the initial choice of the jet opening angle via $\Gamma_{0,\jet} \simeq 1/(1.4 \theta_\jet)$ \citep{mizuta_opening_2013}. 
Combining the hydrodynamic and the electromagnetic tensors, we find the expression for the total stress-energy tensor for a magnetized fluid:
\begin{equation}
    T^{\mu\nu} = \rho_\jet h^* c^2 u^\mu u^\nu + \left( P_\jet + \dfrac{b^2}{2}\right) \eta^{\mu\nu} - b^\mu b^\nu \quad .
\end{equation}
where $h^* = h_\jet(1+\sigma_\jet)$ is the total hydromagnetic enthalpy of the jet, $u^\mu = \Gamma(1,\vec{\beta})$ the proper 4-velocity of the fluid, $P_\jet$ the jet pressure which is related to the jet density via an ideal relativistic fluid's equation of state $P_\jet = \rho_\jet c^2 (h_\jet - 1)/4$. The terms $b^\mu$ and its 4-module $b^2$ are the comoving magnetic fields. 
From the energy-momentum flux term $T^{0\mu}$ of the previous expression (accurately normalized via calculating $\rho_\jet$ self-consistently), we find the total jet luminosity:
\begin{equation}
\label{eq: L_j2}
    L_\jet = 2\pi r^2_\jet \int_0^{\theta_0} \left[ \rho_\jet h^*_\jet  c^2 \Gamma^2_{r,\jet}  \beta_{r,\jet} - B_r (\vec{\beta}\cdot \vec{B}) - \Gamma^2_{r,\jet} (\vec{\beta}\cdot \vec{B})^2 \beta_{r,\jet} \right] c \sin\theta\de\theta \quad .
\end{equation}
with $\Gamma_{r,\jet} = \Gamma_{0,\jet}$.
We inject top-hat jets with constant values across the injection area, low magnetization ($\sigma_\jet \simeq 10^{-2}$) and a mix of toroidal and radial magnetic fields, i.e., $B_r=\chi B_0/\sqrt{1+\chi^2}$ and $B_\phi=B_0/\sqrt{1+\chi^2}$ where $\chi = B_r/B_\phi$.   

\subsection{Shock finder}
To accelerate particles, we need a suitable acceleration site. 
This is found in strong shocks. 
We identify and flag cells as shock zones in our RMHD simulations by imposing the following four conditions that a cell should meet at once: (i) converging flows described by $\nabla\cdot \vec{v} < 0$; (ii) pressure jump between adjacent cells: $|\nabla P |/ \vec{P}_{\min,t} \cdot \Delta \vec{x} \geq \epsilon_\thresh $; (iii) filtering out spurious shock by measuring the misalignment between temperature and density gradients: $\nabla T \cdot \nabla \rho > 0$; (iv) strong relativistic Mach number $\mach_\s>1$. 
We take the pressure jump to be fairly moderate $\epsilon_\thresh  = 3$. $|\nabla P|/\vec{P}_{\min,t}$ represents the ratio between the central difference of the pressure (the scalar product between the second-order accurate pressure gradient and the grid spacing along the direction of the gradient) and the local minimum of the pressure $\vec{P}_{\min,t}$ between adjacent cells along the directions $q_m = [i,j,k]$  \citep{mignone_pluto_2011}.
We measure the Mach number of the shocks $\mach_\s$ by calculating the sound speed $v_\mathrm{sound} = \beta_\sound c$ at least three cells ahead of the shocked cells in the comoving frame, and it is defined as $v_\sound  \equiv \sqrt{\gamma P_\up / (h_\up \rho_\up)}$, 
such that the generalized relativistic Mach number for magnetized fluid is thus defined as the ratio of the shock's 4-velocity in the fluid frame $u_\sh$ and the sonic 4-velocity in the shock's near upstream $u_\mathrm{ms}$, which reads $\mach_\s= u_\sh/u_\mathrm{ms} $.

\section{Preliminary results}
\label{sec: results}

\begin{figure}
    \centering
    \includegraphics[width=1\linewidth]{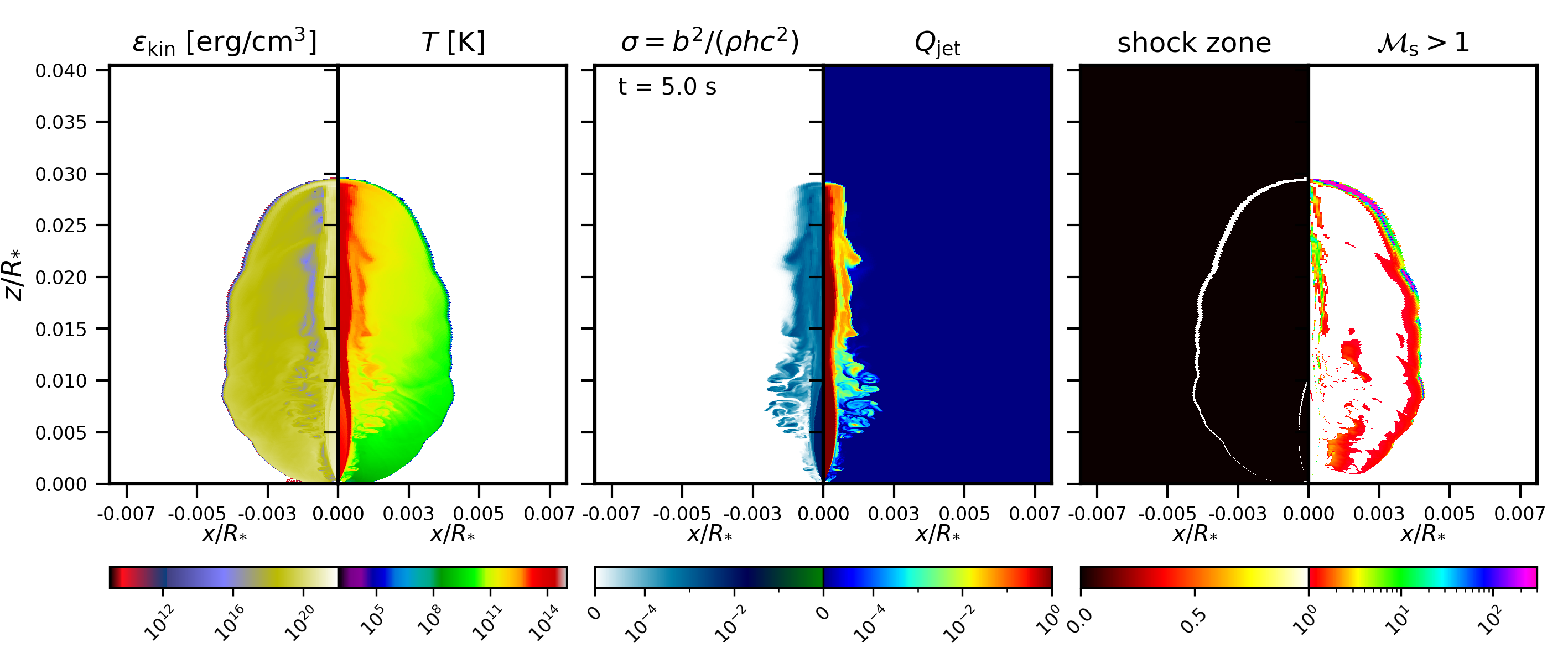}
    \includegraphics[width=1\linewidth]{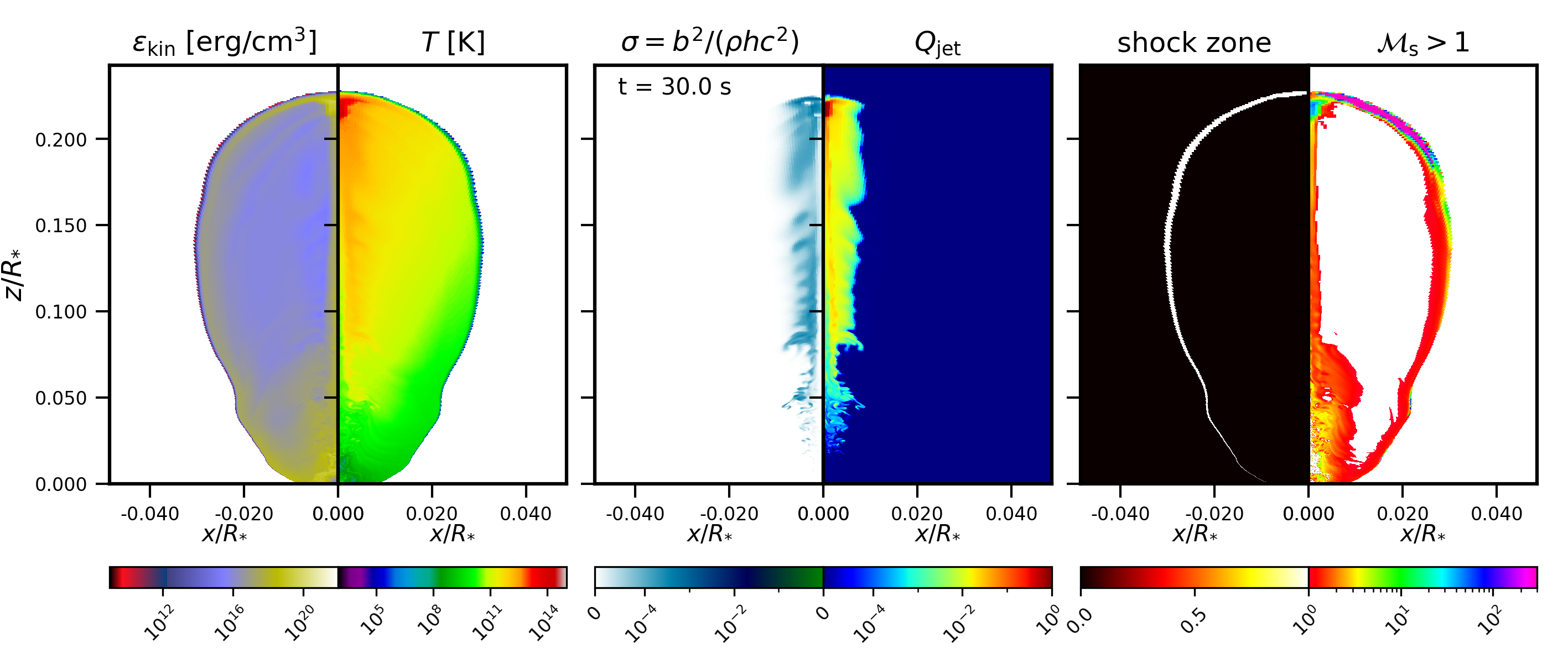}
    \caption{Snapshots of the RMHD simulation of a jet which is drilling its way out of the stellar atmosphere of a BSG star, with $L=10^{51} \ergsec$, initial Lorentz factor of $\Gamma_0=7$, and negligible magnetization $\sigma=0.01$, at two different times. For each row we show, in the following order: the kinetic energy density, the temperature, the relativistic magnetization, a scalar tracer of the jet advected with the fluid, the shock zones identified by our algorithm, and the Mach number $\mathcal{M}_\mathrm{s} \geq 1$ of the shocked material. We used an uneven ratio for the plot axes to enhance the width of the jet. The axes are expressed in units of the stellar radius.
    \textit{Top row}: the system after 5 seconds of injection (equal to the engine time). \textit{Bottom row}: the system after 30 seconds.}
    \label{fig: jet_sim}
\end{figure}

We performed a detailed RMHD simulation of a powerful relativistic jet propagating through the hydrogen‐rich envelope of a Blue Supergiant (BSG) progenitor, characterized by a radius $R_* = 50\,R_\odot$ and mass $M_* = 15\,M_\odot$.  The jet was injected with a constant luminosity $L_{\rm jet} = 10^{51}\,\mathrm{erg\,s^{-1}}$, an initial Lorentz factor $\Gamma_0 = 7$ (corresponding roughly to an opening angle of $5$ degrees for the collimation shock, \citep{mizuta_collimated_2006}), and a magnetization parameter $\sigma = 10^{-2}$.  
Energy and momentum were uniformly deposited through a narrow ($\theta_0 = 2^\circ$) polar cap, producing a top‐hat jet profile.  
We allowed the central engine to operate for $t_{\rm e}=5~$s via an active inner radial boundary, then switched it off and continued the simulation for an additional $25~$s to follow the subsequent evolution.

Figure~\ref{fig: jet_sim} displays snapshots of the simulation at two representative times: immediately after the engine turns off ($t = 5\,$s) and at a later stage ($t = 30\,$s).  
In each panel, we show, from left to right, the pressure distribution, the jet tracer indicating mixing between the jet and the stellar material, the cells flagged as shocks by our algorithm, and the local relativistic Mach number $\mathcal{M}_\mathrm{s} >1$. 
The axes are scaled in units of the stellar radius $R_*$.
During the active phase, the jet head advances rapidly into the stratified envelope, reaching approximately $0.03\,R_*$ by $t = 5\,$s.  
A high‐pressure cocoon forms around the jet head as shocked stellar and jet material are redirected laterally; by this time, the cocoon has expanded to nearly $0.004\,R_*$ in the transverse direction.
Two distinct shock regions are clearly visible: the external cocoon shock, which drives into the ambient envelope, and the termination shock at the jet head, where the relativistic outflow abruptly decelerates.  
Additionally, an internal collimation shock arises within the cocoon.  
This feature is created when the over‐pressurized cocoon re-collimates the jet stream, producing a strong oblique shock inside the jet channel itself.

After the engine is switched off at $t_\engine$, the jet tail, which travels nearly at the speed of light, remains causally disconnected from the head due to the finite jet length at shutdown.  
Consequently, the jet head continues its unimpeded forward motion for several seconds, and no choking occurs within the simulated time frame, leaving only an axial cylindrical high-velocity structure which shrinks over time (the relativistic ``bullet'').  
As the cocoon and jet head propagate into regions of decreasing density, the external cocoon shock accelerates, increasing its average Mach number.  
In contrast, the collimation shock fades away soon after engine turn‐off, as the internal pressure gradient relaxes and the cocoon expansion becomes more isotropic.  By $t = 30\,$s, only the cocoon‐driven forward shock and the residual termination shock remain as prominent sites of energy dissipation.

These high‐Mach, high‐Lorentz shocks may provide conditions for particle acceleration.  
To quantify the non-thermal output of the system, we extract fluid variables (density, pressure, magnetic field strength, and flow velocity) within each shock‐flagged cell. 
We can distinguish two distinct regimes: i) the injection regime, where the jet core (collimation) shock and the jet's forward and cocoon's shock work are strong shocks; ii) the post-injection regime, where the collimation shock disappears, leaving behind only the cocoon shock, which continuously accelerates inside the star as it goes down the density gradient.

Overall, our results demonstrate that for a BSG progenitor with $R_* = 50\,R_\odot$, $M_* = 15\,M_\odot$, and a jet of luminosity $10^{51}\,\rm erg\,s^{-1}$, and engine time $t_\engine = 5~\s$, the jet chokes in about a minute after the launch. The double‐shock structure and evolving Mach numbers indicate that strong shocks are developed both at the jet core and in the external cocoon shock.  The disappearance of the collimation shock after engine shutdown shifts the dominant acceleration sites outward, suggesting that any emergent neutrino or electromagnetic signatures will originate primarily from the forward‐propagating shocks after the first tens of seconds.  These findings provide a foundation for our forthcoming work on particle acceleration and multi‐messenger predictions, which will enable us to derive spectra and light curves of reprocessed photons and neutrinos. Both these signatures will be crucial for defining observational strategies for detecting choked jet phenomena in CCSNe, as preliminarily investigated in \cite{fasano2021} and \cite{zegarelli_towards_2024}. \\
We will extend our parameter study to encompass both Blue and Red Supergiant progenitors, exploring three distinct jet luminosities and two different initial Lorentz factors. By varying these properties and considering a range of observer viewing angles, we aim to identify the combinations that maximize the detectability of high‑energy neutrino emission from gamma‑ray–suppressed, choked‑GRB events (\cite{pais_neutrino_radiation_prep_2025}, \textit{in prep.}).

\vspace{0mm}

\begingroup
\footnotesize
\setlength{\bibsep}{1pt}
\bibliographystyle{ICRC}
\bibliography{references}
\endgroup

\clearpage

\end{document}